\documentstyle{mn}
\input{psfig}
\begin{document}
\newcommand{\bvec}[1] {{\bf {#1}}}
\newcommand{\pdv}[2]{\frac{{\rm\partial} #1}{{\rm\partial} #2}}
\newcommand{\fdv}[2]{\frac{{\rm d} #1}{{\rm d} #2}}
\newcommand{\idv}[1]{{\rm d}#1}
\newcommand{\eng}[2]{#1 \times 10^{#2}}
\newcommand{\half}{\frac{1}{2}}
\newcommand{\omb}{\Omega_{\rm b}}
\newcommand{\seu}{M\,d^2\,\omb^3}
\bibliographystyle{plain}
\title[Simulations of superhumps and superoutbursts]
{Simulations of superhumps and superoutbursts}
\author[J.~R. Murray]{J.~R. Murray\thanks{Now at The Astrophysical Theory Centre, 
Australian National University, ACT 0200, Australia}\\
Canadian Institute for Theoretical Astrophysics,
University of Toronto,
Ontario, M5S 3H8,
Canada\\
}
\maketitle
\begin{abstract}
We numerically study the tidal instability of accretion discs in close
binary systems using a two-dimensional SPH code. We 
find that the precession rate of
tidally unstable, eccentric discs does not only depend upon the binary mass ratio $q$.
Although the (prograde) disc precession
rate increases with the strength of the tidal potential, we find that
increasing the shear viscosity $\nu$ also has a significant prograde
effect. Increasing the disc temperature has a retrograde impact upon
the precession rate.

We find that motion relative to the binary potential results in
superhump-like, periodic luminosity variations  in  the outer reaches of
an eccentric disc. The nature and location of the luminosity
modulation is a function of $\nu$. Light curves most similar to
observations are obtained for  $\nu$ values appropriate for a dwarf nova in outburst.

We investigate the thermal-tidal instability model for
superoutburst. A dwarf nova outburst is simulated by instantaneously
increasing $\nu$, which causes a rapid radial expansion of the
disc. Should the disc encounter the $3:1$ eccentric inner Lindblad
resonance and become tidally unstable, then tidal torques become much
more efficient at removing angular momentum from the disc. The disc
then shrinks and $\dot M_{\rm d}$ increases. The resulting increase in
disc luminosity is found to be 
consistent with the excess luminosity of a superoutburst.

\end{abstract}

\section{Introduction}
In a previous paper (Murray 1996, hereafter Paper I) we described a
two dimensional  
smoothed particle hydrodynamics (SPH) code specifically designed for
thin disc problems.
We showed that
the standard SPH artificial viscosity term
introduces a kinematic shear viscosity, $\nu$, into a disc simulation that
can be accurately estimated analytically 
by taking the SPH equation of motion to the
continuum limit. For a two dimensional SPH code with the standard
cubic spline kernel one obtains
\begin{equation}
\nu = \frac{1}{8}\,\zeta\,c\,h.
\label{eqn:SPHvic}
\end{equation}
where $\zeta$ is the coefficient of the linear artificial viscosity
term, $c$ is the sound speed, and $h$ is the SPH smoothing length
(which has been kept fixed in all calculations described herein).
Equation~\ref{eqn:SPHvic} was verified with axisymmetric
ring-spreading calculations, and by comparing the stationary end state
of a simulation of a disc in a binary with axisymmetric theory.

Two simulations of discs in low mass ratio $(q=3/17)$ binary systems
evolving under steady mass transfer from the secondary were
described. In agreement with calculations in Whitehurst
(1988)\nocite{whitehursta} and  Hirose \& Osaki
(1990)\nocite{hiroseandosaki}, the discs encountered the binary
system's $3:1$ eccentric inner Lindblad resonance (see Lubow
1991a\nocite{Lubowtheory}). As a result, both discs became
considerably  eccentric and, as seen in the inertial frame, precessed
slowly in the prograde direction. Whitehurst (1988) proposed that the
superhump phenomenon seen in some short period cataclysmic variables
was the signature of a precessing, eccentric disc. Indeed,
in the simulations described in Paper I
approximately ten to twenty per cent of
the total  luminosity from the eccentric disc was
modulated at the frequency of the motion of the disc's semi-major axis
in the binary potential.

In this paper we describe a more thorough numerical investigation of
the so called {\it tidal instability} of accretion discs in close
binary systems. Simulations for a range of binary mass ratios, disc
temperatures, and shear viscosities are described in order to
illustrate the relative importance of these parameters in determining
whether discs are tidally unstable. Eccentricity growth rates and disc
precession rates are compared with observations and with analytical
results obtained by Lubow
(1991a, 1992)\nocite{Lubowtheory, Lubowthird}.
We go on to demonstrate that tidal instability can not only explain
the superhumps seen in SU~UMa systems, but also provide a long-term
increase in energy dissipation in the disc that is consistent with the
excess luminosity of SU~UMa superoutbursts. Simulations described
here provide considerable support for the thermal-tidal instability
model for superoutburst proposed by Osaki (1989)\nocite{Osaki89}.

In section 2 we briefly outline our current understanding of SU~UMa
systems, superhumps and superoutbursts. In section 3.1 we discuss our
numerical scheme, focusing on alternative methods of introducing shear
viscosity into disc calculations. Then in section 3.2 we detail the
assumptions and parameter settings used in the various
simulations. Section 4 is devoted to numerical results of superhump
simulations (4.1), and superoutburst simulations (4.2). 
Our major conclusions are summarised in section 5.

\section{Preamble}
\subsection{The SU~Ursae Majoris systems}
Cataclysmic variable (CV) stars are semi-detached binary systems,
with a white dwarf as the detached component (primary), and a low mass
star as the contact component (secondary). For a comprehensive review
of the observations and theory of CVs see
Warner~(1995)\nocite{Warnerbook}. Typically a few tenths of a solar
mass, the secondary star loses mass to the more massive 
white dwarf via Roche lobe
overflow. 
In the absence of strong magnetic fields, the transferred gas forms an
accretion disc around the white dwarf.
CVs display a wide variety of
eruptive behaviours, and non-magnetic systems are phenomenologically
categorised on this basis.

Dwarf novae are CVs that undergo luminosity increases (outbursts), 
typically of 2--5 magnitudes,
which last a few days and recur on a time-scale of days to
weeks. The SU~Ursae Majoris (SU~UMa) systems form a subclass of dwarf
novae that also exhibit 
a related but distinct phenomenon, called 
superoutburst. 
These eruptions are approximately one magnitude brighter than normal
outbursts and last as long as a couple of few weeks rather than just a
few days.
The accretion disc has been found to be the source of the 
excess luminosity of both normal outbursts and superoutbursts.
Because superoutbursts often very closely follow a
normal outburst, and because of similarities in the rises to maximum,
it is likely that superoutbursts are triggered by normal outbursts.

For a recent review
of the observational details of the superoutburst phenomenon, the
reader is once again referred to Warner (1995). Here we
merely highlight a few relevant points. 
Firstly, it appears that all SU~UMa systems
have short periods and that all dwarf novae with short periods are
SU~UMa systems. Thorstensen et al. (1996)\nocite{thorstensen} tabulated
the orbital periods for 26 SU~UMa systems.
Of these, TU~Men had the longest orbital period with  $P_{\rm
b}=169.34\,{\rm min}$.
Even TU~Men's period is  short relative to the orbital periods
of ordinary dwarf novae.
Warner (1995) made
the remark that all dwarf novae with periods $P_{\rm b} \la 3\,{\rm hrs}$
are eventually identified as  SU~UMa systems.  

The shorter a CV's orbital period is, the smaller its
binary mass ratio, $q={M_{\rm sec}}/{M_{\rm pri}}$, is expected to be. 
Those
dwarf novae with periods less than three hours (i.e. the SU~UMa
systems) are thought to have $q \la 0.3$.

\subsection{Superhumps}
Superhumps are large amplitude modulations of a CV's optical
light curve.
Typically, the superhump period, $P_{\rm sh}$, is a few per cent longer
than the binary period.
Until recently, superhumps had been seen only  in
SU~UMa systems during their superoutburst phase. This is no longer the
case.
Patterson et al. (1995)\nocite{patterson95} observed the  superhumps in the
SU~UMa system V1159 Orionis  to persist well after
superoutburst had ended.
Modulations tentatively identified as superhumps have also
been observed
in several  other short period  cataclysmic variables (see
Patterson \& Richman 1991,
Retter, Leibowitz \& Ofek 1997\nocite{pattersonrichman,retter}) that don't exhibit
classic dwarf novae activity. 

Thorstensen et al. plotted the fractional superhump period
excess for  26 SU~UMa systems
as a function of $P_{\rm b}$ (their figure 9), and obtained a line
of best fit

\begin{equation}
\frac{P_{\rm sh}-P_{\rm b}}{P_{\rm b}} = -0.0344+(0.0382\,{\rm
hr}^{-1})\,P_{\rm b}.
\label{eq:thorstensen}
\end{equation}
With $P_{\rm b}$ thought to be an increasing function of $q$, we can
infer that the superhump period is also dependent upon the binary mass
ratio. Significantly, the superhump period excess does not go to zero
with the binary period. This empirical relationship suggests that in
the limit of very small $q$, the superhump period would be shorter than
the binary period.

For a given system $P_{\rm sh}$  is commonly found not to be
completely stable but may in fact decrease slightly over the course of
a single superoutburst (see e.g. Patterson et al. 1995). 
 
\subsection{The Thermal-Tidal Instability Model for Superoutbursts}
Osaki (1989) proposed a model for SU~UMa superoutbursts that
incorporates the thermal instability model for normal dwarf nova
outbursts, and a tidal instability that induces the accretion
disc to become eccentric. The details of the thermal instability
model can be found in Cannizzo (1993a)\nocite{cann1}. 
Central to the model is the hypothesis that angular momentum transport
is much more efficient on the hot branch of the limit-cycle than on
the cold branch. A dwarf nova outburst then is a period of enhanced
mass flux $\dot M_{\rm d}$ through the disc. 
Cannizzo (1993b\nocite{cann2}) found that a disc with
Shakura-Sunyaev viscosity parameters
$\alpha_{\rm hot}=0.2$ and $\alpha_{\rm cold}=0.01$ best reproduced
the observed dwarf nova outbursts of SS Cygni.

The tidal instability is due to the $3:1$ eccentric inner Lindblad
resonance (see Lubow 1991a\nocite{Lubowtheory}). Only in systems with
small binary mass ratio $(q \la 0.3)$ can the disc extend to the
resonant radius, and then only if angular momentum transport in the
disc is reasonably efficient. It should only be possible to excite the
resonance in short period CVs that are on the hot branch (whether
permanently or temporarily) of the thermal limit-cycle.

In Osaki's so called thermal-tidal instability model for
superoutbursts, each normal outburst causes a net radial expansion of
the disc. When the disc expands sufficiently it encounters the
resonance and becomes eccentric. Osaki proposed that 
in its eccentric state, the disc would be subject
to greatly enhanced tidal torques that increased the mass flux, $\dot
M_{\rm d}$, through the disc and therefore increased the
disc luminosity, explaining why superoutbursts are significantly
brighter than normal outbursts.
 In the inertial frame, the eccentric disc exhibits a
slow prograde precession, and superhumps are explained as a beat
phenomenon between the motion of the disc and the motion of the
binary.
The thermal-tidal instability model is reviewed in Osaki (1996).

\section{Numerical Method}
\subsection{SPH}
The two-dimensional smoothed particle hydrodynamics (SPH) code used
here has already been described in Paper I. In that paper, we 
demonstrated the code's
ability to reliably model an accretion disc with a given
 kinematic shear viscosity, $\nu$.
By taking the equations of motion for the SPH particles to the
continuum limit, one finds that
the standard SPH artificial viscosity term
generates a force per unit mass
\begin{equation}
\bvec{a}_{\rm v}=\frac{\zeta h \kappa}{2\Sigma}\,
(\bvec{\nabla}\cdot(c \Sigma \bvec{S})+\bvec{\nabla}
(c \Sigma \bvec{\nabla} \cdot \bvec{v})).
\label{eq:genvisc}
\end{equation}
Here $\zeta$ is the coefficient of the linear SPH artificial viscosity
term, $h$ is the
smoothing length, $\Sigma$ is the disc  column density, 
$\bvec{v}$ is the velocity, and 
\begin{equation}
S_{ab}=\pdv{v_a}{x_b}+\pdv{v_b}{x_a}
\end{equation}
is the deformation tensor. In two dimensions
\begin{equation}
\kappa=\frac{\pi}{4}\,\int_0^{\infty} {r^2}\,\fdv{W}{r}\, dr.
\end{equation}
For the standard cubic spline kernel,
$\kappa=\frac{1}{4}$. 
If the sound speed is only slowly varying in space, the SPH artificial
viscosity produces a shear viscosity with magnitude given by
equation~\ref{eqn:SPHvic}.
In regions where the dilatation is non-zero, a bulk viscosity is also
produced. 

In the inner accretion disc, orbits are approximately circular and the
shear term dominates the dissipation. In the outer disc where the
tidal influence of the secondary is stronger, the bulk term is more
important. The radial extent of the disc depends upon the relative
magnitudes of the shear and bulk viscosities (Papaloizou and Pringle
1977)\nocite{papapringle}). $r_{\rm m}$, the mean radius of the largest
simply periodic test particle orbit that doesn't intersect other such
orbits of smaller radius (Paczy\'nski 1977\nocite{paczynski}), is a simple estimator of
the boundary between shear and bulk viscosity dominated regions of the disc.

The standard linear SPH artificial viscosity term 
includes shear and bulk
viscosity in a fixed ratio.
One can derive SPH interpolant estimates of $\nabla^2\,\bvec{v}$, and
hence include independent shear and bulk viscosity terms in the SPH
equations.
Two possible SPH formulations were introduced by Flebbe et al. (1994)\nocite{flebbeandall} and Watkins et
al. (1996)\nocite{steveandall}. However these methods require two
levels of interpolation to generate the second derivatives of the
velocity and so smooth information over a larger region.
Furthermore, in these formulations, the 
forces between two particles are not antisymmetric and along the
particles' line of centres. As a result, angular momentum is only
conserved at a global level to an accuracy dictated by the interpolation scheme.
As we wish to use the dissipation
terms to transport angular momentum it is highly desirable that
angular momentum
be conserved as precisely as possible. 
With the dissipation term we use, linear
and angular momentum are conserved to machine accuracy at the particle level.

In standard SPH, there is a `viscous switch' that sets the viscous
force between two particles to zero if they are receding from one
another. This is to ensure that dissipation only occurs in regions of
compression. For our disc SPH code, the switch is disabled. The
quadratic term in the standard SPH artificial viscosity formulation is
not used. This term always produces a repulsive force between two
particles and so causes problems at the outer edge of the disc where
particles are flung to very large radii. 

\subsection{Simulation Details}
We use the scalings of Lin and Pringle
(1976)\nocite{linandpringle}, where the total system mass $M$, the
binary orbital angular velocity $\Omega_{\rm b}$, and the interstellar
separation $d$, are all set to one. 

We approximate the gravitational potential of the binary to be that of
two point masses in circular orbit of radius $d$ about each other. We
have an open inner boundary condition, i.e. there is a hole in the
computational domain of radius $r_{\rm wd}$ centred on the position of the
primary (point mass). At the end of each time step particles lying
within this circle are removed from the simulation. In all simulations
described here we use $r_{\rm wd}=0.02 \,d$. Particles that end a time step
either within the secondary star's Roche lobe or a distance $r > 0.9
\, d$ from the primary's centre of mass are also deleted. 

An isothermal equation of state (i.e. constant sound speed c) is
used. We also set the SPH artificial viscosity parameter $\zeta$ to
be constant, so that the kinematic and bulk viscosities will
be constant. As in Paper I, energy dissipation in the disc is
calculated using the SPH energy equation (see their equation 17).

Mass is added to the calculation in one of two ways. If we wish to
incorporate the mass transfer stream in the simulation we add a single
particle per time step $\Delta t$
at the inner Lagrangian point. In principle, the
stream boundary conditions at $L_1$ are a function of the binary mass
ratio $q$ (see Lubow and Shu 1975\nocite{lubowandshu}). In practice we
always added particles with an initial speed $v_{\rm
inj}=0.1\,d\,\omb$, in a direction $0.367$ radians prograde of the binary axis.
If we do not wish to include the mass transfer stream in the
calculation we simply add one particle per time interval $\Delta t$ in
a circular Keplerian orbit of radius $r_{\rm c}$. Standard practice is
to choose $r_{\rm c}$ to be the circularisation radius, which is again
a function of the mass ratio. For these simulations, when we did add
particles in circular orbit, it was at $r_{\rm c}= 0.1781\,d$. Unless
specifically stated, we always chose the interval between particle
addition to be $\Delta t = 0.01 \,\omb^{-1}$

As we have an isothermal equation of state and constant $\nu$, the fluid equations are
independent of the mass scaling. In other words our choice of
particle mass  has no effect upon the simulation outcome beyond
scaling the surface density. In
every simulation then,
we chose the particle mass to give us a 
mass transfer rate from the secondary star of
$10^{-9}\,M_\odot {\rm yr}^{-1}$ assuming 
$P_{\rm b}=0.063121$ days (the period for OY Car).

\section{Results}
\subsection{Tidal Instability and Superhumps}
We have completed twelve simulations in which we began with zero
initial mass and followed the viscous evolution of the disc under
steady mass addition. The results are summarised in Table~\ref{tab:results}.
Note that
runs 4 and 5 were  previously described in Section 4 of
Paper I as simulations 1 and 2 respectively. 
In run 6, the shear viscosity  
$\nu = \eng{5}{-5} d^2\,\omb$.
In runs 7,8 and 12, $\nu = \eng{2.5}{-5} d^2\,\omb$.
In all the other calculations, $\nu = \eng{2.5}{-4} d^2\,\omb$.

We followed
each calculation for a  time (see column 4 of the table) that was sufficient for us to
determine whether the disc would become eccentric or not, and also
long enough for us to accurately measure precession periods. 
For example, as it was apparent that run 1 had reached an essentially
steady state, and as it was clear that the disc would not become
eccentric,  the simulation was terminated at $t=1652.05 \,\omb^{-1}$.

Computing requirements also
dictated the length of various simulations. The total number of
particles in a disc at steady state is proportional to the rate at
which particles are added to the disc, and inversely proportional to
the viscosity. Of course the computing time increases with the number
of particles. On the other hand, if we wished to follow
the viscous evolution of a relatively inviscid disc, we are required to
follow the calculation for many binary orbital periods. Hence when we
reduced $\nu$ by a factor 10 in runs 7 and 8, we also reduced the rate of
particle addition (but not mass addition)  to the disc to one
sixth the standard value.

Most of the simulations result in eccentric, precessing discs. Our
technique for determining the period of the disc's motion in the
binary frame $P_{\rm d}$, differs from the method used in Paper
I. There we measured the motion of the disc's centre of mass and
obtained a mean value for $P_{\rm d}$. We estimated an uncertainty in
our figure from the standard deviation of the data.
However, it is possible to identify $P_{\rm d}$ in simulation
`luminosity' data as well.
At regular time intervals ($0.01 \,\omb^{-1}$)
we record the rate at which energy is being dissipated in various
regions of the disc, to obtain time series
that can be analysed for periodicities in the same manner as
observational data. We loosely equate the energy dissipation rate to
luminosity. These time series, which we shall refer to as `simulation
light curves', are illustrated in figure~\ref{fig:diss} and discussed
later in this section.

\begin{figure}
\mbox{\psfig{figure=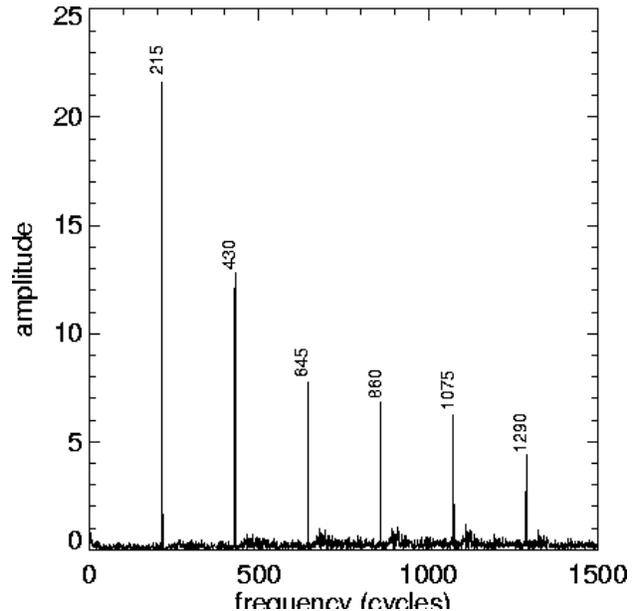,height=8cm}}
\caption{Fourier transform of the outer disc $(r
> 0.20 \, d)$ light curve from run 10. The  time
series covers a $1400.00 \,\omb^{-1}$ time period with a resolution of
$0.01 \,\omb^{-1}$. We thus  estimate $P_{\rm d} = 1400.00/215\,\omb^{-1}$.}
\label{fig:pwspec}
\end{figure}

To obtain a mean period $P_{\rm d}$ we simply Fourier transform the
simulation light curve. As noted in Paper I, simulation light curves
from the entire disc are made quite noisy by the action of a small
number of particles releasing large amounts of energy in the inner
disc. Thus we used light curves obtained from disc regions at radii 
$r > 0.20\,d$. The Fourier spectrum from run 10 is shown in 
Fig.~\ref{fig:pwspec}. Clearly there is only one frequency in the
data at 215 cycles per $1400.00 \,\omb^{-1}$. The other peaks are
simply higher harmonics. The spectra from the other runs are similarly
unambiguous, with the peaks being only one or two cycles wide. The
values  given for ${P_{\rm d}}/{P_{\rm b}}$ for runs 4 and 5
are consistent with, but
more precise than, the values given in Paper I.

\begin{table*}
\begin{minipage}{175mm}
\caption{Summary of simulations and their basic results}
\label{tab:results}
\begin{tabular}{ccccrcccccr}
Run & $q$ & $c$ & $\zeta$ & Time
& ${P_{\rm d}}/{P_{\rm orb}}$ & ${P_{\rm d}}/{P_{\rm orb}}$ 
& $\lambda_{\rm t}$ & $\lambda_{\rm m}$ & final $S_{(1,0)}$ &
$t_{\rm inst}$\\

& & $d\,\omb$ & & $\omb^{-1}$ &(mean)&(max)& $ \omb$ &$ \omb$ & & $\omb^{-1}$ \\

1 & 0.29 & 0.02 & 10.00 & 1652.05 & no oscillation & --- & --- & ---
& $0.001\pm0.001$ & --- \\

2 & 0.25 & 0.02 & 10.00 & 2000.00 & $1.128 \pm 0.004$ & ${}^e$ &
$0.26\pm 0.01$ & $0.012\pm 0.001$ & $0.15\pm 0.01$ & 729.81\\

3 & 0.25 & 0.06 & 3.333 & 2000.00 & $1.073 \pm 0.003$ & $1.099
\pm 0.006$ & 0.3 & $0.018 \pm 0.001$ & $0.53 \pm 0.01$ & 181.24\\

4 & 3/17 & 0.02 & 10.00 & 1000.00 & $1.079 \pm 0.009$
& ${}^f$ & $0.10 \pm 0.01$ & $0.046 \pm 0.002$ & $0.31 \pm 0.02$ & 367.88\\

5 & ${3/17}^a$ 
& 0.02 & 10.00 & 1000.00 & $1.074 \pm 0.009$
& ${}^f$ &${}^g$ & $0.058 \pm 0.002$ & $0.35 \pm 0.03$ & 63.66\\

6 & ${3/17}^a$ & 0.02 & 2.000 & 870.20 & $1.064 \pm 0.001$
 & $1.074\pm 0.001$ & ${}^g$ & $0.052 \pm 0.004$ & ${0.220}^h$ & 183.75\\

7 & ${3/17}^b$ & 0.02 & 1.000 & 6374.76 & $1.049 \pm 0.001$
& $1.070 \pm 0.001$ & ${}^g$ & $0.020 \pm 0.004$ & $0.27 \pm 0.01$ & 689.28\\

8 & ${3/17}^{a,b}$ & 0.02 & 1.000 & 8551.08 & $1.047 \pm 0.001$ &
$1.069 \pm 0.002$ & ${}^g$ & $0.025 \pm 0.003$ & $0.29 \pm 0.01$ & 480.66 \\

9 & 1/9 & 0.02 & 10.00 & 1618.77 & $1.053 \pm 0.003$ & ${}^f$ & 0.059
-- 0.066 & $0.036 \pm 0.002$ & $0.39 \pm 0.05$ & 198.59\\

10 & 1/9 & 0.06 & 3.333 & 2000.00 & $1.036 \pm 0.002$ &
$1.051 \pm 0.004$ & 0.06 -- 0.07 & $0.0190 \pm 0.0005$ & $0.52 \pm 0.02$ &181.24\\

11 & 1/19 & 0.02 &  10.00 & 1436.20 & no oscillation & --- &
--- & --- & $0.001\pm 0.001$ & ---\\

12 & ${1/19}^a$ & 0.02 & 1.000 &1000.00 & $1.018 \pm 0.001$ & ${}^f$ 
& 0.0027 & $0.018\pm0.002$ & ${0.232}^h$ &342.19\\

\\
\end{tabular}
${}^a$ mass added in circular orbit at radius $r_{\rm c}$.\\
${}^b$ particles added singly every $\Delta t=0.06\,\omb^{-1}$\\
${}^e$ period is stable to within uncertainty in measurement of
${P_{\rm d}}/{P_{\rm orb}}$.\\
${}^f$ ${P_{\rm d}}$ increases over first 10 or so periods.\\
${}^g$ use the same value as in run 4.\\
${}^h$ eccentricity increasing at a rate $0.0001$ per scaled time unit.
\end{minipage}
\end{table*}

The mean values of ${P_{\rm d}}/{P_{\rm b}}$ are listed in column
6 of Table~\ref{tab:results}. As noted in Paper I (see their Table 3),
for a given mass ratio $q$, different authors have obtained a range of
values for ${P_{\rm d}}/{P_{\rm b}}$. 
One can append  the results of three
dimensional simulations completed by Whitehurst
(1994) to that table.
Because everybody had used
their own numerical scheme, it was  not clear whether the differences
were physical or numerical.

From Table I however it is now apparent that  ${P_{\rm d}}/{P_{\rm
b}}$ is not simply a function of $q$. It is also influenced by the
temperature and dissipation in the disc. For example the only
difference between runs 9 and 10 is a factor 3 increase in the
pressure. Note that $\zeta$ was reduced a factor three from run 9 to run 10 so 
that only the pressure and not the viscosity differed between the two 
calculations. The mass ratio and mode of mass addition are also
identical. Both discs evolved to eccentric, precessing final
states. Yet the hotter disc (run 10) had a significantly shorter
$P_{\rm d}$. This is consistent with the analysis of Lubow (1992), who found that gas
pressure had a retrograde influence on the disc precession. The same
comparison can be made between runs 2 and 3. Because the sound speed in run 3 is 
three times greater than in run 2 but the viscosity parameter $\zeta$ is three  times smaller, 
the two simulations differ only in the value of the pressure. We note however that the 
disc in run 2 was
only ever barely cognizant of the resonance. 
$P_{\rm d}/P_{\rm b}$ is very large in run 2.
The greater pressure in
run 3 on the other hand allowed the disc to expand further outwards
and interact more strongly with the resonance. 

In Paper I we found there to be no evidence of any trends, secular or
periodic, in $P_{\rm d}$. Now, with a larger set of simulations, and
an improvement in our analysis, we must revise that statement. We
borrow from the observers' bag of tricks and use O--C diagrams to
detect changes in $P_{\rm d}$. We take a simulation light
curve, smooth it somewhat, and 
obtain timings for maxima in the
disc dissipation. These are our `observed' maxima. The test period in
each case is
the mean value of $P_{\rm d}$, given in column 6 of Table~\ref{tab:results}.
We have now observed two types of change in $P_{\rm d}$ that occur as the
disc evolves from an axisymmetric state to a non-axisymmetric steady state.

\begin{figure}
\mbox{\psfig{figure=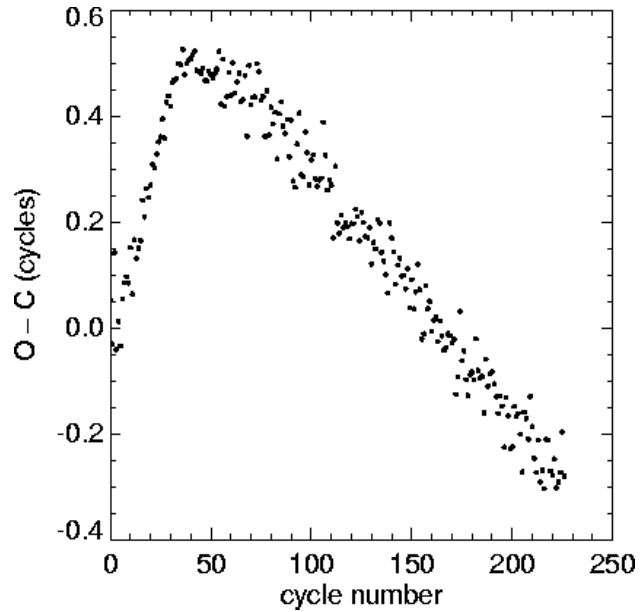,height=8cm}}
\caption{O--C diagram for run 3 relative to a test period of $6.774
\,\omb^{-1}$. We take the dissipation rate `light curve' from the
outer disc ($r > 0.2\,d$) and smooth it by taking a running mean over
200 points. The first `observed' peak in the light curve occurred at
$t = 450.72\,\omb^{-1}$. The first forty or so cycles have mean period
$6.905 \pm 0.038\,\omb^{-1}$.  }

\label{fig:oc3}
\end{figure}

The O--C diagram for run 3 is shown in Fig.~\ref{fig:oc3}. For the first 40 or
so revolutions  of the disc in the binary frame, the observed $P_{\rm
d}$ has a value that is slightly longer than the test
period. At this time the disc eccentricity is growing
exponentially. Then, at about the same time as the eccentricity growth
slows, $P_{\rm d}$ begins to decrease. $P_{\rm d}$ 
and the eccentricity then stabilise together.
We only observed this behaviour in those discs that became eccentric
reasonably slowly. 
We used the O--C diagrams to estimate the
initial value for $P_{\rm d}/P_{\rm b}$ (listed in Table~\ref{tab:results}).

\begin{figure}
\mbox{\psfig{figure=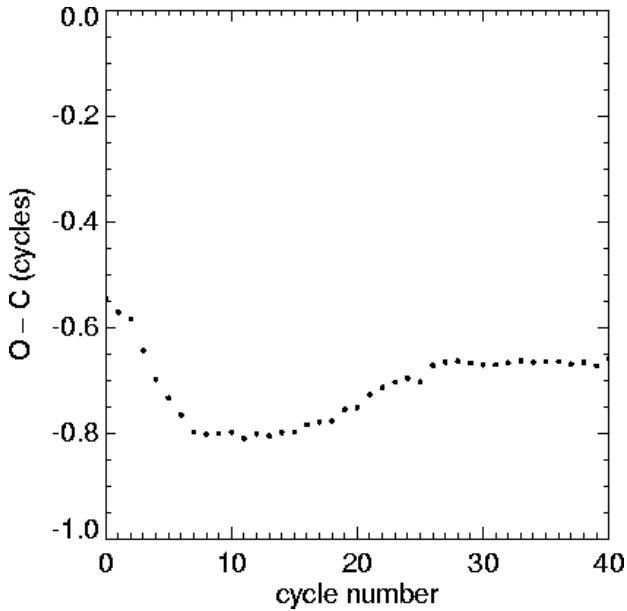,height=8cm}}
\caption{First 40 cycles of the O--C diagram for run 6 
relative to a test period of $6.736
\,\omb^{-1}$. We take the dissipation rate `light curve' from the
outer disc ($r > 0.2\,d$) and smooth it by taking a running mean over
200 points. The first `observed' peak in the light curve occurred at
$t = 155.50\,\omb^{-1}$. 
}
\label{fig:oc6}
\end{figure}

When the disc initially encounters the resonance, the oscillations
induced in the simulation `light curve' are of comparable magnitude to
the background noise. There is thus considerable variation in the
length of the first few cycles. 
Nevertheless it is apparent that
the mean period of the first few
cycles is less than the test period i.e. the first few
cycles are shorter than the long term average period.
Fig.~\ref{fig:oc6}, the O--C diagram of run 6 shows this clearly. The
initial 25-30 cycles roughly form a concave upwards parabola
indicating that $P_{\rm d}$ is increasing. Then $P_{\rm d}$ apparently
decreases again to reach a stable value at about the same time that
the eccentric instability saturates.
Simulations where this is seen are marked in Table~\ref{tab:results}
with an $f$.

The last four columns of Table~\ref{tab:results} summarise the
behaviour of the eccentric mode. 
As in Lubow (1991b)~\nocite{LubowSPH} and in Paper I,
we determine the
strength of the $(l\,\theta -m\,\omb t)$
 mode, $S_{(l,m)}$, by Fourier decomposing the
disc density distribution in azimuth and time. In particular we
use $S_{(1,0)}$ to estimate the disc eccentricity.
Column 8 gives a theoretical estimate for the exponential growth rate
of the eccentricity in each disc. Column 9 lists the measured  growth
rate of $S_{(1,0)}$.
Column  10 gives the strength of the eccentric mode at the end of each
simulation. This value indicates how strongly the disc is affected by
the resonance. The final column, which lists the time at which
$S_{(1,0)}= 0.01$, is a measure of when the disc initially encounters
the resonance.

The theoretical estimates for the growth rate, $\lambda_{\rm t}$, are
obtained using the analysis in Lubow (1991a)\nocite{Lubowtheory}.
There, an expression was obtained for the eccentricity growth rate,
$\lambda_0$, of a narrow ring of material at the resonance  radius
$r_{\rm res}$. The growth rate for a disc
\begin{equation}
\lambda_{\rm t} = C \lambda_0,
\label{eq:lambdao}
\end{equation}
where the correction factor
\begin{equation}
C = \frac{M_{\rm res}\,\,e_{\rm res}}{M_{\rm d}\,{<e>}}.
\label{eq:correction}
\end{equation}
Here $M_{\rm res}$ is the mass of material in the resonance region,
$M_{\rm d}$ is the total disc mass, $e_{\rm res}$ is the eccentricity
at the resonance radius, and $<e>$ is the mass averaged eccentricity
of the entire disc.  We see from equation~\ref{eq:correction} that if
there is sufficient mass at the resonance, and the eccentricity there
is larger than the mass averaged value, it is possible for the
eccentricity of a disc to grow at a faster rate than an `ideal' narrow ring.
We point this out as the rise times of superoutbursts are relatively
short, which implies that the disc must become eccentric reasonably rapidly.

\begin{figure}
\mbox{\psfig{figure=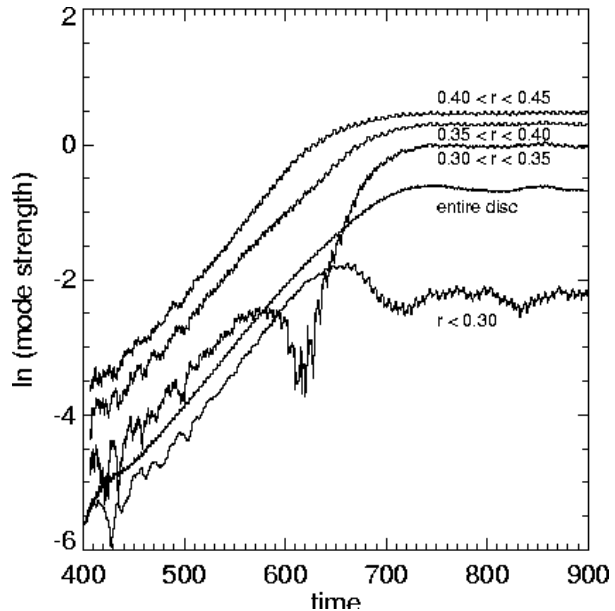,height=8cm}}
\caption{Logarithm of the eccentric mode strength $S_{(1,0)}$
as a function of
time for various regions of the disc in run 3. Time units are $\omb^{-1}$.}
\label{fig:eccrad}
\end{figure}

\begin{figure}
\mbox{\psfig{figure=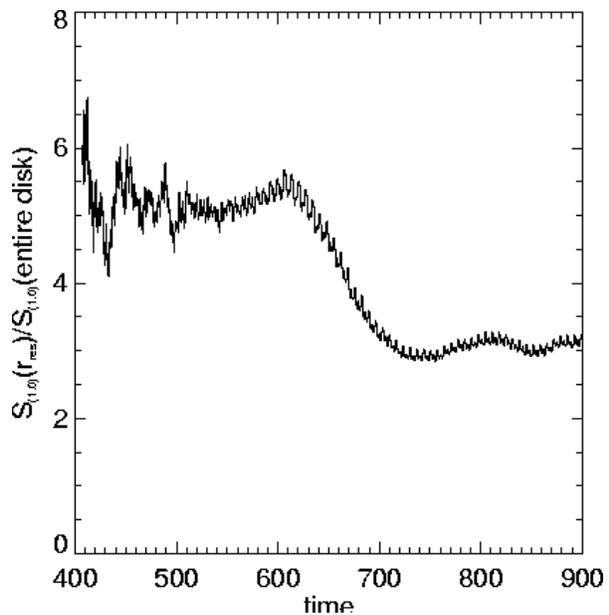,height=8cm}}
\caption{Ratio of the strength of the eccentric mode in the annulus
$0.40\,d<r<0.45\,d$ to $S_{(1,0)}$ for the entire disc.}
\label{fig:eonebar}
\end{figure}

\begin{figure*}
\mbox{\psfig{figure=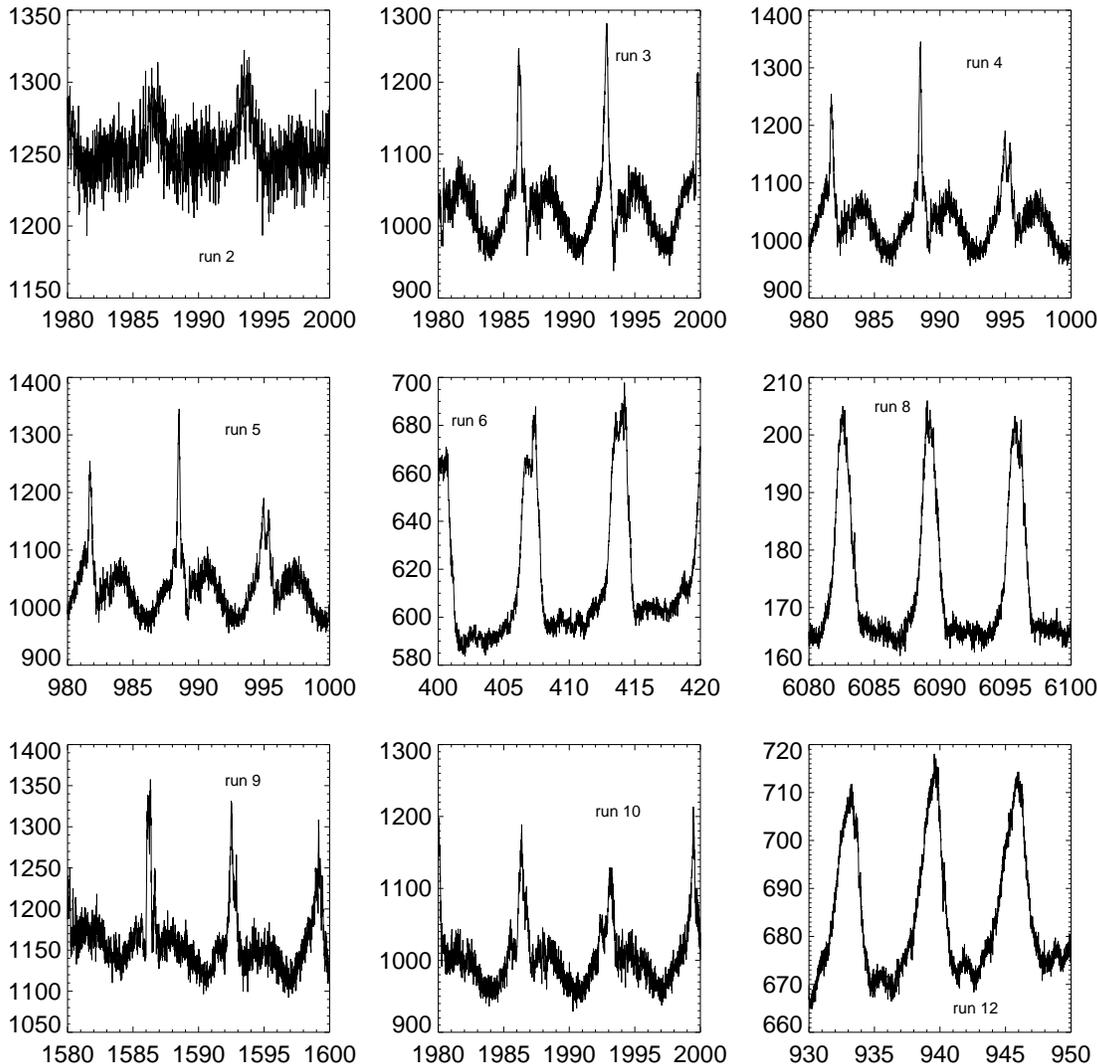,width=15cm}}
\caption{
Each frame shows a few cycles of the light curve of one of the runs
once the disc precession has fully developed. In each case we plot the
rate of energy dissipation ($\seu$) at radii $r > 0.05\,d$ as a function of time $(\omb^{-1})$.
In runs 6 and 12 the
disc has not reached a stationary state as the total mass (and the
background dissipation rate) is still increasing. In all other cases
the disc is in a `steady state'.}
\label{fig:diss}
\end{figure*}

In calculating the values for $\lambda_{\rm t}$ in Table~\ref{tab:results}, and
also in Paper I, we assumed that the eccentricity of the disc was
uniform. 
Figure~\ref{fig:eccrad}, a plot of $S_{(1,0)}$ at different radii
as a function of time, shows that the eccentric
mode strength in run 3 is an increasing function of the radius. 
We see from figure~\ref{fig:eonebar}
that $S_{(1,0)}$ at the resonant radius is approximately three to
five times the value for the entire disc. It is not unreasonable to assume
the eccentricity behaved similarly in the other simulations (with the
possible exception of run 2 where the resonance only weakly interacted
with the disc). Therefore the values in column 8 may underestimate the
ideal growth rate of the eccentricity by a factor five or so. This
would make the disagreement between  $\lambda_{\rm t}$ and
$\lambda_{\rm m}$ more pronounced, with theory and simulation then
only being in reasonable agreement for run 12. 
In all other cases the measured
eccentricity growth is much slower than predicted by analysis.

SPH simulations by Lubow (1991b, 1992) that included only the $m=3$
component of the tidal potential (i.e. that component directly
involved in the resonance), produced eccentricity growth  rates that
were in good agreement with the theory. However, when the full binary
potential was used in the calculation, the eccentricity growth rate
was much reduced. Lubow (1992) found that non-resonant stresses excited
in the disc by the $m=2$ component of the tidal field were the
principal culprits. 

Figures \ref{fig:eccrad} and \ref{fig:eonebar} show the eccentricity
growing equally rapidly at different radii. The time at which the mode
strength reaches a plateau however, is a function of the radius. There is a
huge inexplicable dip in the eccentric mode strength of the annulus $0.30\,d < r <
0.35\,d$. In the absence of this trough, the mode appears to be growing
at the same rate in this annulus as it is at other radii. 

The periodic stressing of the disc as it rotates in the binary frame
introduces a cyclic component into the simulation light curves,
which we shall refer to as the `superhump signal'. The `superhumps'
from nine of the simulations are collected in Figure~\ref{fig:diss}.
The light curves produced by the low viscosity simulations (runs 6, 8 and 12 are illustrated)
are clearly different from those of the high viscosity
calculations. In the latter we see sharp spikes superposed on a lower
amplitude, noisy signal which has two peaks per cycle. 
The low viscosity runs on the other hand have one
tall reasonable broad peak per cycle, superposed on an almost flat
background luminosity. The high frequency noise is much less apparent in these
curves. We would also like to single out the light curve of simulation
2. Of all the high viscosity simulations it is the only one without
the characteristic sharp spikes. The remaining superhump signal is of
small amplitude and is almost swamped by high frequency noise.
This is consistent with our
interpretation that the disc in run 2 only barely encountered the
resonance, and didn't interact strongly with it.

In Paper I we showed that the sharp spikes in the high viscosity
simulations were due to dissipation in small knots of material. These
knots formed near apastron on highly eccentric orbits in the outer
disc. The dissipation occurred when the knots, moving towards
periastron, collided with gas on less eccentric orbits in the inner disc.
These knotty structures are far less prominent in the low viscosity
simulations. In general, the outer boundaries of the low viscosity discs
are smoother and more regular than those of the high viscosity discs (see
figures 8 and 12 of Paper I for density plots of runs 4 and 5), and
the low viscosity discs have a less fragmentary appearance.

Recall from Paper I that the Shakura-Sunyaev viscosity parameter
\begin{equation}
\alpha(r)=\frac{1}{8}\,\zeta\,h\,\Omega(r)\,c^{-1},
\label{eq:equivalpha}
\end{equation}
 where $\Omega(r)$ is the angular velocity at radius $r$ in the disc.
Thus $\alpha(r_{\rm res})=3/16$ in simulations 7,8 and 12, $3/8$ in
run 6, and $15/8$ in the other runs.
The lower values are quite close to observational estimates of
$\alpha$ for dwarf novae in superoutburst. This is pleasing because it
is precisely these simulations that give results most similar to the
observations.

\begin{figure}
\mbox{\psfig{figure=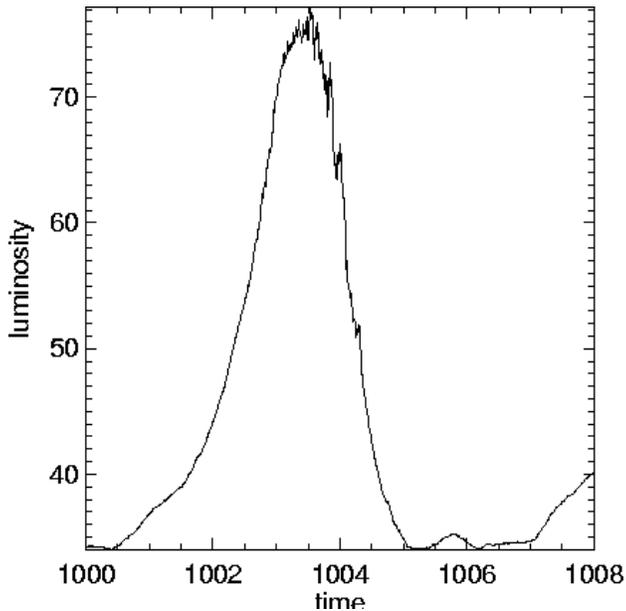,height=8cm}}
\caption{Luminosity, i.e. energy dissipation rate, ($\seu$) from radii $r >
0.30\,d$ as a function of time ($\omb^{-1}$) for simulation
12.
}
\label{fig:mu95diss}
\end{figure}

\begin{figure*}
\mbox{\psfig{figure=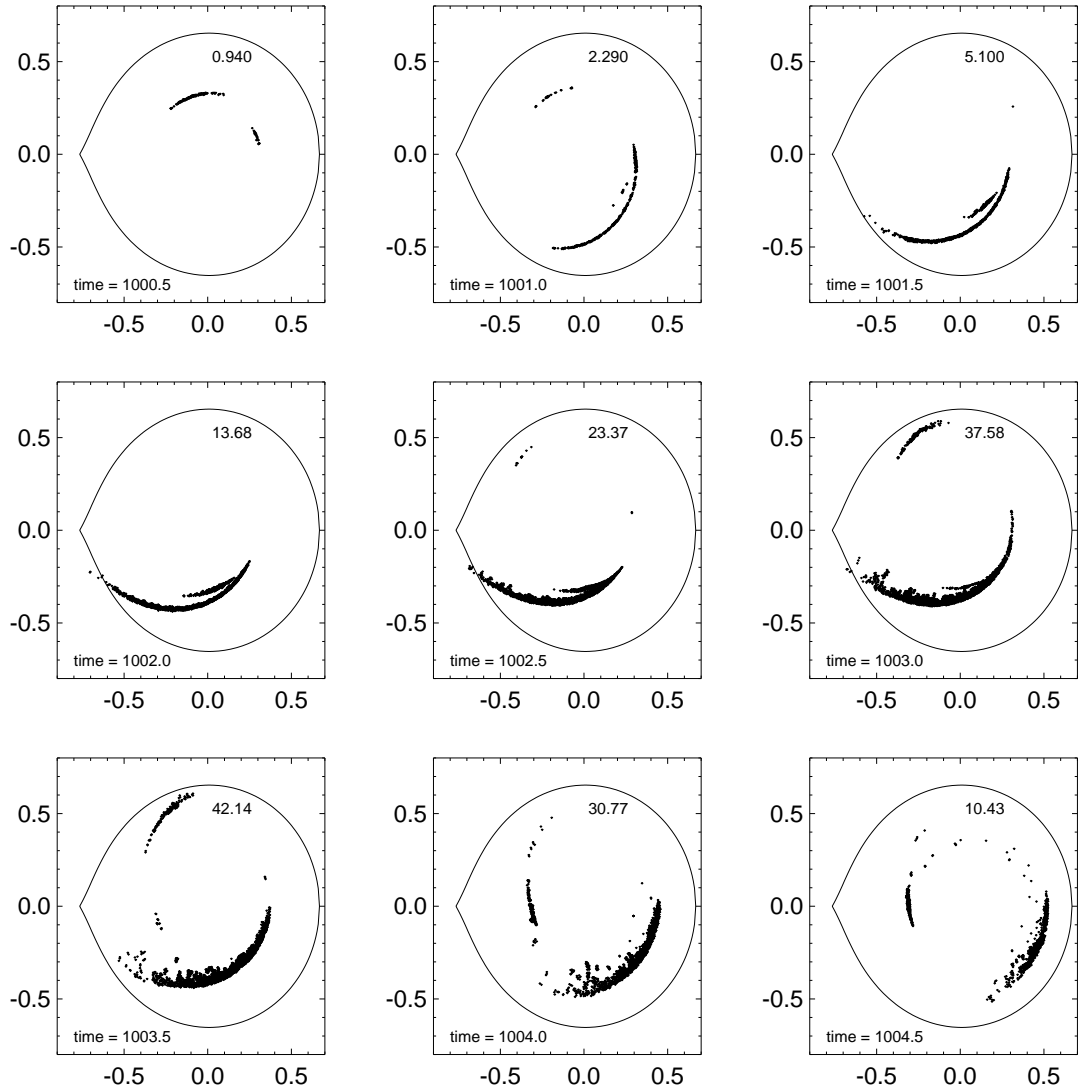,width=15cm}}
\caption{The `superhump source' for simulation 12. 
These nine plots, covering most of a disc rotation period $P_{\rm d}$,
map the disc regions that are responsible for the signal
shown in figure~\ref{fig:mu95diss}. 
In each frame, on axes corotating with the binary,
we plot the `most luminous' particles in the outer disc, i.e.
those particles at radii $r > 0.30\,d$ that have
a luminosity greater than $\eng{5.0}{-3}\seu$. The total luminosity due to
the selected particles is shown in the upper right of each frame, and
the time is printed in the lower left. 
The x and y axes are scaled to
the interstellar separation $d$. The Roche lobe of the primary is
drawn in each frame. The system is rotating anticlockwise. The white
dwarf primary is located at the origin.
}
\label{fig:shsignal}
\end{figure*}

In our simulations, the superhump signal is produced in the disc's
outer regions. In figure~\ref{fig:mu95diss} we show approximately one
superhump cycle from simulation 12, plotting the energy dissipated at
radii $r > 0.3\,d$ as a function of time. Compare the superhump's
amplitude of approximately $42\,\,\seu$ with the steady background
luminosity at these radii  of only $34\,\,\seu$ (keeping in mind that
the disc is not in steady state). 
In figure~\ref{fig:shsignal} we
show the locations of the particles responsible for the superhump
signal. At each time shown we selected all the particles in the outer
disc ($r > 0.3\,d$) with a luminosity greater than a set 
threshold $(\eng{5.0}{-3}\,\,\seu)$,
and plotted them. The total luminosity of the selected particles is
shown in the upper right corner of each frame. 
Matching these values to figure~\ref{fig:mu95diss} it is clear that we
have isolated the source of the hump in the simulation light curves.

\subsection{Simulations of superoutburst}
In this section we describe two simulations that test a key assumption
of the thermal-tidal instability model; that tidal forces remove
angular momentum more efficiently from a resonant disc than from a
non-resonant disc. One can calculate the
rate of angular momentum removal from a narrow gas ring at the $3:1$
resonance (see equation 62 of Lubow 1991a). But how efficiently can angular
momentum be extracted from a disc that has just expanded into the
resonance region at the beginning of a (normal) outburst? Is the
resultant increase in energy dissipation large enough to account for
observed superoutburst luminosities? 

To simulate an outburst, we instantaneously increase the artificial
viscosity in a simulation by a factor 5 or 10. This gives us only a crude
approximation of an outburst, with the entire disc being suddenly forced to
change from a cool, inviscid state to a hot, viscous state.
A more detailed outburst model, that allows different parts of the disc to
change state at different times could be easily implemented by allowing
each SPH particle to have its own $\zeta$.  

For the first calculation we subjected the low viscosity, $q=3/17$
disc (run 6) to a five-fold increase in viscosity, by setting  $\zeta=10.0$ at
time $t=100.00 \,\omb^{-1}$. All other
parameters were left unchanged. For reasons that will become clear
shortly, we shall refer to this simulation as the `superoutburst calculation'.
In the absence of any change
in the viscosity, the run 6 disc encountered the resonance and became
tidally unstable at $t_{\rm inst}=183.75$. However, with the aid of
the sudden increase in the viscosity, the `superoutburst' disc expanded and became
resonant almost immediately.

The  strongest non-resonant tidal
response in the disc occurs in the $(2,2)$ mode. 
As such, the  $(2,2)$ mode strength  provides a measure of the radial
extent of the disc. 
The relative strengths of the $(1,0)$ and $(2,2)$ modes
also tell us whether the disc is  resonant or
non-resonant, or equivalently whether the disc is stationary in the
inertial frame or
stationary in the binary frame. 

\begin{figure}
\mbox{\psfig{figure=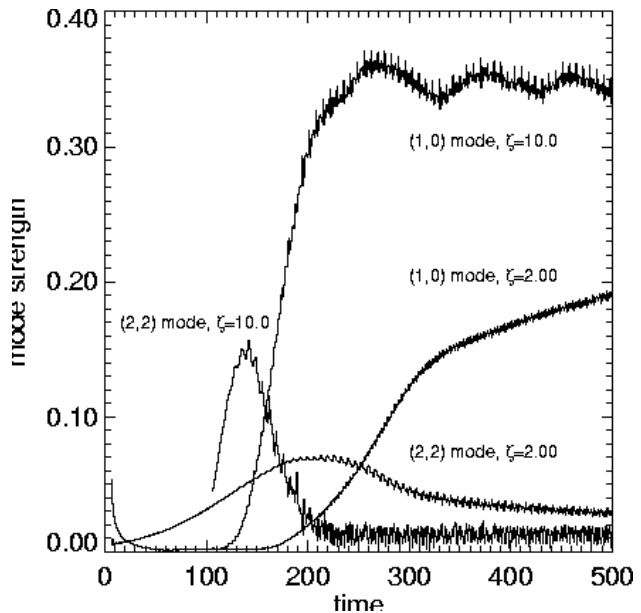,height=8cm}}
\caption{
Strengths of the eccentric $(1,0)$ and the tidal $(2,2)$ modes for the
superoutburst simulation ($\zeta=10.0$) and for
run 6 ($\zeta=2.00$). Time (in $\omb^{-1}$) is measured from the
beginning of run 6. Thus the superoutburst simulation starts at
$t=100.00 \,\omb^{-1}$.
}
\label{fig:supermodes}
\end{figure}

In  figure~\ref{fig:supermodes}, the
$(1,0)$ and $(2,2)$ modes for the superoutburst calculation and for
run 6 are plotted as functions of time. The sharp initial increase in
the $(2,2)$ mode strength shows that the superoutburst disc
rapidly expanded in response to the increase in viscosity. However,
tidal instability then set in. As the disc eccentricity rose,
tidal torques became more efficient at removing angular momentum from
the disc. Consequently the disc shrank once more, as witnessed
by the subsequent decline in the $(2,2)$ mode. The weakening of the
$(2,2)$ mode implied an increase in the inward mass flux
$\dot M_{\rm d}$ which, as we shall see below, caused an increase in
the disc luminosity.

Tidal instability also caused the disc in run 6 to shrink, but much
less dramatically.
Note that the eccentricity in run 6 continued increasing even
after the tidal instability had became
saturated at $t=300\,\omb^{-1}$. In this case, both the total disc mass and angular
momentum continued to grow because tidal torques were still not strong
enough to remove all the  angular momentum being added with new disc
material at $r_{\rm c}$.

\begin{figure}
\mbox{\psfig{figure=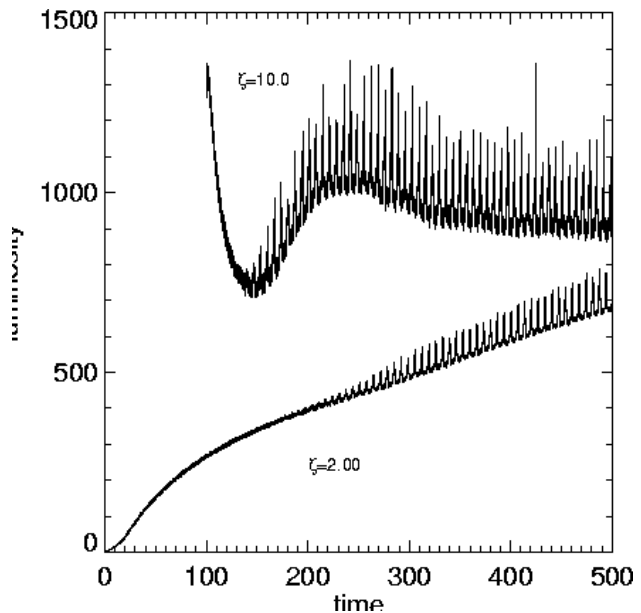,height=8cm}}
\caption{Luminosity $(\seu)$ from radii $r > 0.05\,d$ in the
superoutburst simulation. The equivalent light curve for
 run 6 is also plotted (lower curve). Note that a
superhump signal is apparent in both curves.}
\label{fig:compdiss}
\end{figure}

Light curves for run 6 and the superoutburst simulation are plotted in
figure~\ref{fig:compdiss}.
Note that the superoutburst disc was initially much more luminous than
the disc in run 6. However the excess luminosity decayed fairly
rapidly. Recall that the inwards mass flux in an axisymmetric,
Keplerian disc is given by
\begin{equation}
\dot M_{\rm d}=6 \pi r^\half \pdv{}{r}(\nu \Sigma r^\half).
\label{eq:keplerflux}
\end{equation}
From equation~\ref{eq:keplerflux} it is apparent that
the instantaneous viscosity increase caused $\dot M_{\rm d}$ to
increase to a rate  greater than the mass transfer rate from
the secondary. This lead to a general decline in density in the disc,
which in turn caused
$\dot M_{\rm d}$ to decay to a steady state value. Thus the initial
decline in the superoutburst disc luminosity is explained. 

But the luminosity of the superoutburst simulation did not then 
flatten out. Rather, it reached a minimum value of $707.46\,\seu$ at
$t=148.92\,\omb^{-1}$ and then rose once more. The minimum in the
luminosity coincided almost exactly with the maximum
strength of the $(2,2)$ mode which occurred at
$t=141.83\,\omb^{-1}$. 
The mean luminosity in the superoutburst simulation 
peaked at about $1000\,\seu$ at time
$t=240\,\omb^{-1}$. Thus there was a 40--50 \% increase in the energy
output of the disc 
over a time interval of about 15 orbital periods. In other words the
disc brightened by about 0.4 magnitudes over approximately one day,
which is consistent with observations of the rise to superoutburst
maximum.

The disc luminosity then slowly declined as the disc adjusted to the
increased tidal torques at the outer edge. The light curve is
qualitatively similar to the extended decline of a
superoutburst. The limitations of our simulation
probably preclude closer comparison with observations. For a start we
used an isothermal equation of state (i.e. the speed of sound was
constant). The effective Shakura-Sunyaev viscosity parameters of our
simulations were much larger than those found by Cannizzo (1993b) to be
appropriate for a normal outburst. For the superoutburst simulation we
had $\alpha_{\rm hot}(r_{\rm res})=15/8$ and $\alpha_{\rm
cold}(r_{\rm res})=3/8$. 
Also, the ratio $\alpha_{\rm hot}/\alpha_{\rm cold}=5$ in our
simulation, which was rather small. Furthermore,
rather than having a constant
$\alpha$ throughout, our simulations have constant kinematic viscosity
$\nu$. One would expect that the proportion of the disc mass that
enters the resonance region, and hence the superoutburst luminosity
will be dependent upon $\nu$.

Note also in figure~\ref{fig:compdiss} that the superhump signal
appeared well before the `superoutburst maximum'. Superhumps are not
usually observed until approximately a day after superoutburst maximum
has occurred. However, recent observations of both ER~UMa (Kato,
Nogami \& Masuda 1996\nocite{kato}) and V1159~Orionis 
(Patterson et al. 1995\nocite{patterson95}) 
found superhumps on the rise to superoutburst maximum. 

\begin{figure}
\mbox{\psfig{figure=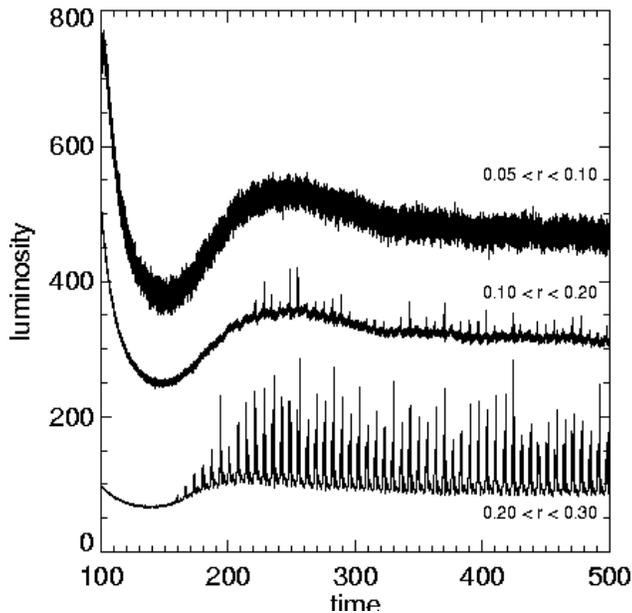,height=8cm}}
\caption{Luminosity $(\seu)$ of various disc annuli in the
superoutburst simulation. Notice that the superhump signal
predominately comes from radii $r > 0.20\,d$. From bottom to top, the
curves have minima at $t=137.37$,$143.03$ and $154.58\,\,\omb^{-1}$.
}
\label{fig:suumadiss}
\end{figure}

Some idea of the radial dependence of the superoutburst luminosity can
be gained from figure~\ref{fig:suumadiss}, in which light curves
generated by different annuli of the superoutburst disc are plotted.
The initial luminosity dip up to time $t \simeq 200\,\omb^{-1}$ was
shallower at larger radii. Indeed, at the largest radii (not shown)
the dip was in fact a hump, reflecting the initial expansion of the
disc in response to the increase in
$\zeta$. Figure~\ref{fig:suumadiss} shows that the superoutburst in
the inner disc is delayed somewhat with respect to the superoutburst
in the outer regions. This is consistent with there being an
enhancement in the mass flux at the outer edge of the disc which then
works its way inwards. As a result the superoutburst light became
somewhat bluer with time. Also notice that the superhump signal first became
apparent, and had the largest amplitude, in the light curve of the
annulus $0.20\,d<r<0.30\,d$. A small superhump signal appeared somewhat
later in the $0.10\,d<r<0.20\,d$ light curve.

\begin{figure}
\mbox{\psfig{figure=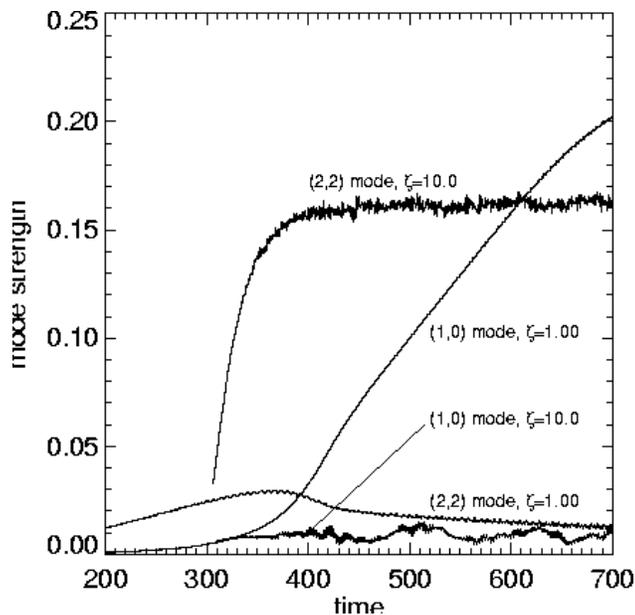,height=8cm}}
\caption{
Mode strengths for the normal outburst simulation ($\zeta=10.0$), 
and for  run 12
($\zeta=1.00$). Time (in $\omb^{-1}$) is measured from the
beginning of run 12. Thus the normal outburst simulation starts at
$t=300.00 \,\omb^{-1}$.
}
\label{fig:dudmodes}
\end{figure}

We completed a second `normal outburst' calculation, using run 12 ($q=1/19$,
$\zeta=1.00$) as the seed simulation. At $t=300.00\,\omb^{-1}$ we
reset the viscosity to $\zeta=10.0$, but left the other parameters
unchanged. The $(1,0)$ and $(2,2)$ mode strengths for the normal outburst
simulation and run 12 are plotted as functions of time in
figure~\ref{fig:dudmodes}. Shortly after $t=300.00\,\omb^{-1}$ the
radial expansion of the run 12 disc was arrested as the disc became
unstable ($t_{\rm inst}=342.19$). However at mass ratio $q=1/19$ the
$3:1$ resonance is too weak to drive a disc with $\zeta=10.0$ to tidal
instability (see run 11). Hence, in the normal outburst simulation, the sudden
increase in $\zeta$ simply caused the disc to expand radially. The
eccentricity in the normal outburst simulation did not increase. 

\begin{figure}
\mbox{\psfig{figure=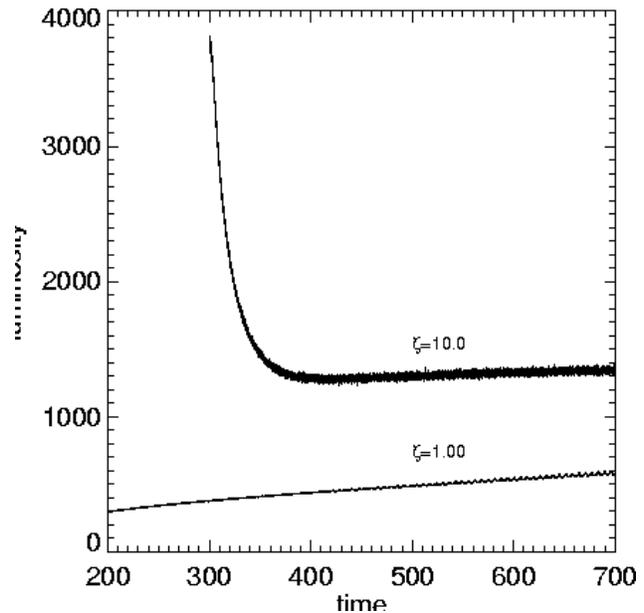,height=8cm}}
\caption{
Luminosity $(\seu)$ from radii $r > 0.05\,d$ in the normal outburst
simulation. The equivalent light curve for run
12, is also plotted (lower curve). Note that a superhump signal is
only apparent for run 12.
}
\label{fig:duddiss}
\end{figure}

Light curves for run 12 and the normal outburst simulation are shown in
figure~\ref{fig:duddiss}. The normal outburst simulation luminosity declined
rapidly from a high initial value as the disc adjusted to the
increased viscosity. This time however, without a tidal instability
sponsored increase in the mass flux, there was no second rise in the
luminosity. The diminutive superhumps of run 12 clearly show how weak
the $3:1$ resonance is for this mass ratio.

The simulations described in this section provide support for the
thermal-tidal instability model for superoutburst. In particular, we
find that when a rapidly expanding disc encounters the $3:1$ eccentric
inner Lindblad resonance and becomes tidally unstable, large
quantities of angular momentum are removed from the disc causing it to
shrink once more. The energy released is sufficient to account for the
excess luminosity of a superoutburst.

\section{Conclusions}
\begin{figure}
\mbox{\psfig{figure=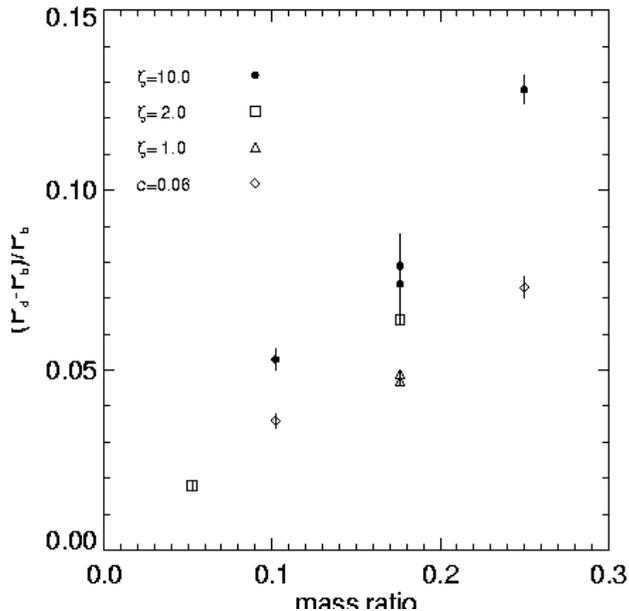,height=8cm}}
\caption{Superhump period excess  (with error bars shown) as a function of binary mass ratio
for the  simulations of section 4.1. }
\label{fig:precess}
\end{figure}

We conclude that the disc precession period (which we identify with
the superhump period) is {\bf not} uniquely determined
by the
binary mass ratio. This point is clearly illustrated by
figure~\ref{fig:precess}, where the `superhump period excesses`, 
$(P_{\rm d}-P_{\rm b})/P_{\rm b}$, for the
simulations of section 4.1 are plotted against $q$. We find that gas
pressure and viscosity are also important in determining the disc
precession period. This conclusion contradicts the findings of
Whitehurst \& King (1991)\nocite{whitehurstandking}. Concluding that
the binary mass ratio was the sole factor that determined the
precession period, they estimated $P_{\rm d}$ from the orbital periods
of  non-interacting test particles moving in the binary's
gravitational potential. Specifically, they estimated that the
superhump period was equal to the period of doubly periodic test
particle orbits in the vicinity of the resonance, 
{\em as measured in the inertial frame}. Setting to
one side for a moment concerns about using non-interacting particles,
in the observations and in the simulations, the superhump phenomenon 
recurs at a common position in the {\em binary frame}. Hence their
estimate should have been 
\begin{equation}
P_{\rm d}\simeq P_{\rm b} (1 + w/\Omega_{\rm b})
\label{eq:whiteking}
\end{equation}
where $\omega$ is the angular rate of precession of the particle's
line of apsides.

We do not wish to overstate the case as
figure~\ref{fig:precess} clearly shows the superhump period excess
increasing with mass ratio, indicating that the tidal potential is
most important in the determination of $P_{\rm sh}$.
However, if we are to use an expression similar to
equation~\ref{eq:whiteking} to predict the binary mass
ratio from the orbital and superhump periods, then viscosity and gas
pressure must be accounted for. We mentioned in section 4.1 that
increasing the gas pressure had a retrograde effect upon the disc
precession. Such an effect may be significant in actual SU~UMa
systems.
Recall that equation~\ref{eq:thorstensen}, the empirical relationship between the superhump and
binary periods, suggests that
systems with very small  orbital period 
(and hence by inference a very small mass ratio) will have $P_{\rm sh}
< P_{\rm b}$.
We suggest that this is
due to gas pressure slowing the disc precession in the manner
described in Lubow (1992)\nocite{Lubowthird}.

We found small but significant changes in the disc precession
period.
In several cases, $P_{\rm d}$ decreased slightly once the resonance
had saturated and the disc eccentricity had stabilised.
Thus it may be possible to use
 the small
changes in $P_{\rm sh}$ over the course of a single superoutburst to 
determine whether disc eccentricity is growing or declining.

Both in Paper I and here (see figure~\ref{fig:shsignal}) we showed
where in the disc the periodic component of the thermal energy
generation occurs.
However, we neglected the details of
the disc's vertical structure in our calculations, and so we must be
very cautious about comparing these results with observational
attempts to isolate the superhump source (e.g. Warner
\& O'Donoghue 1988\nocite{WarnerO'Donoghue}, and O'Donoghue
1990\nocite{O'Donoghue}). 
In other words, the time
taken for energy released on the disc midplane to reach the disc
surface is likely of the order of a dynamical time (${\Omega(r)}^{-1}$).

Billington et al. (1996) found dips in the ultraviolet light from the
inner disc of OY~Car in superoutburst that coincided with the
superhumps. Using a modified eclipse mapping technique, they showed
that the dips could be explained if the disc had a raised outer rim
that obscured the inner disc. Figures \ref{fig:mu95diss} and
\ref{fig:shsignal} showed that in simulation 12 the `superhump source'
accounted for just over  half the outer disc luminosity. This
indicates that there are sufficiently large {\em azimuthal}
temperature gradients in the outer disc to generate the 10-20\%
variations in the height of the outer disc that the observations of
Billington et al. imply. They suggest that the superhump luminosity is
actually ultraviolet radiation from the inner disc that is reprocessed
down to the optical in the raised disc rim. In our simulations, the
periodic component of the `light curve' was of sufficient amplitude to
be comparable to superhumps. We propose that the original superhumps
are powered simply by the variable thermal energy dissipation in the
outer disc. This would naturally generate an outer disc that varied in
height with azimuth. It is possible that reprocessing of light from
the inner disc causes further evolution in the disc rim, leading to
the late superhumps that are seen towards the end of a superoutburst. 

Whitehurst \& King concluded that the growth rate of the tidal
instability was too slow to be consistent with observations of
superoutbursts. They suggested that a burst of mass transfer from the
secondary was required to accelerate the rate at which the disc
becomes eccentric. We do not agree with that conclusion. Our results
show that a disc can become eccentric on a time scale consistent with
the rise time of a superoutburst. Recall that in
figure~\ref{fig:suumadiss} superhump-like modulations of the light curve
were visible on the rise to maximum luminosity. In this simulation
mass was added in circular orbit at radius $r_{\rm c}$. We found that
mass addition from $L_1$ slowed rather than accelerated growth in the
$(1,0)$ mode (not however to a rate that was inconsistent with
superoutburst observations).  Lubow (1994)~\nocite{Lubowfourth} found analytically that the
impact of the stream upon the disc acted to reduce disc eccentricity.

Finally, the simulations of section 4.2 provide strong support for the
thermal-tidal instability model for superoutburst. We approximated a
dwarf nova outburst by instantaneously increasing the shear
viscosity. We found that when a rapidly expanding disc also became
tidally unstable, there was an increase in the disc luminosity that
was consistent with a superoutburst. In our simulation, superhump-like
modulations were apparent on the rise to maximum luminosity. In a
second case, where the expanding disc did not become tidally unstable,
there was no increase in the luminosity. 

The author would like to thank Steve Lubow
for much encouragement,  useful advice and constructive criticism.

\bsp
\end{document}